\documentclass[structabstract]{aa}  
\usepackage{natbib}
 \bibpunct{(}{)}{;}{a}{}{,} 
 
\usepackage{graphicx}

\usepackage{txfonts}
\def\kms{km s$^{-1}$\space}
\def\kmsnospace{km s$^{-1}$}
\def\micron{$\mu$m\space}

\def\deg{$^{\circ}$\space}

\def\h2{H$_2$}

\def\c2{[C\,{\sc ii}]}
\def\13co{$^{13}$CO}
\def\c18o{C$^{18}$O}
\def\12co{$^{12}$CO}

\def\c+{C$^+$}

\def\h2{H$_2$}

\begin{document}

     \title{The scale height of gas traced by [C\,{\sc ii}] in the Galactic plane}
   
\titlerunning{[C\,{\sc ii}] Galactic scale height}
\authorrunning{Langer, Pineda, Velusamy}

   \author{W. D. Langer,
             J. L. Pineda,
           and
              T. Velusamy\thanks{{\it Herschel} is an ESA space observatory with science instruments provided by European-led Principal Investigator consortia and with important participation from NASA.}   
           }
          

   \institute{Jet Propulsion Laboratory, California Institute of Technology,
              4800 Oak Grove Drive, Pasadena, CA 91109-8099, USA\\
              \email{William.Langer@jpl.nasa.gov}
             }

   \date{Received 19 December 2014 / Accepted 10 March 2014}

 

\abstract
{The distribution of various interstellar gas components and the pressure in the interstellar medium (ISM) is a result of the interplay of different dynamical mechanisms and energy sources on the gas in the Milky Way.  The scale heights of the different gas tracers, such as H\,{\sc i} and CO, are a measure of these processes.  The scale height of [C\,{\sc ii}] emission in the Galactic plane is important for understanding those ISM components not traced by CO or H\,{\sc i}.}   {We determine the average distribution of [C\,{\sc ii}]  perpendicular to the plane in the inner Galactic disk  and compare it to the distributions of other key gas tracers, such as CO and H\,{\sc i}.}  {We calculated the vertical, $z$, distribution of [C\,{\sc ii}] in the inner Galactic disk by adopting a model for the emission that combines the latitudinal, $b$, spectrally unresolved BICE survey,  with the spectrally resolved $Herschel$ Galactic plane survey of [C\,{\sc ii}] at $b = 0\degr$.  Our model assumed a Gaussian emissivity distribution  vertical to the plane, and related the distribution in $z$ to that of the latitude $b$ using the spectrally resolved [C\,{\sc ii}] $Herschel$ survey as the boundary solution for the emissivity at $b=0\degr$.} {We find that the distribution of [C\,{\sc ii}] perpendicular to the plane has a full-width half-maximum of 172 pc, larger than that of CO, which averages $\sim$110 pc in the inner Galaxy, but smaller than that of H\,{\sc i}, $\sim$230 pc, and is offset by -28 pc.}  {We explain the difference in distributions of [C\,{\sc ii}], CO, and H\,{\sc i} as due to [C\,{\sc ii}] tracing a mix of ISM components.  Models of hydrostatic equilibrium of clouds  in the disk predict different scale heights, for the same interstellar pressure. The diffuse molecular clouds with [C\,{\sc ii}] but no CO emission likely have a scale height intermediate between  the low density atomic hydrogen  H\,{\sc i} clouds and the dense CO molecular clouds. }

{} 

\maketitle



\section{Introduction}
\label{sec:introduction}

The star formation rate in the Galaxy may be related to the pressure of the interstellar medium (ISM), which itself is a function of the interplay of dynamical processes and energy sources on the interstellar gas.  It has also been suggested that ISM pressure plays a role in the formation of giant molecular clouds \cite{Blitz2004,Blitz2006}.   Thus, the vertical ($z$) distribution of the various ISM gas components is an important parameter for understanding these dynamical processes throughout the Milky Way. In the Galaxy, interstellar clouds are distributed in a thin disk about the mid-plane at $b=0\degr$ and their scale height depends, in hydrostatic equilibrium, on a number of factors, including thermal pressure, the random motion of the clouds, magnetic pressure, ionization pressure, and the gravitational force of the stars and gas in the disk.  Thus, determining the $z$--distribution, which may be different for various ISM components, provides information about these parameters. The scale height of the diffuse atomic hydrogen clouds is known from extensive maps of the H\,{\sc i} 21-cm line \cite[c.f.][]{Boulares1990,Dickey1990}  and that for the dense molecular hydrogen clouds from large scale  maps of the $^{12}$CO $J=1\rightarrow 0$ rotational line \cite[e.g.][]{Sanders1984,Dame1987,Bronfman1988,Clemens1988,Malhotra1994,Dame2001,Jackson2006}. The scale height for H\,{\sc i} and H$_2$ (as traced by CO) clouds are  different and each varies by a factor of $\sim$3 across the inner Galaxy, and increases in the outer Galaxy \citep{Narayan2002}.  In addition to tracing gas that can be observed in H\,{\sc i} and CO, the 1.9 THz emission from ionized carbon, [C\,{\sc ii}], traces the H$_2$ gas where carbon is ionized but little, or no, CO or neutral carbon is found (the CO-dark H$_2$ gas), and also traces the warm ionized medium (WIM).

The scale height for clouds traced by [C\,{\sc ii}] is not well established because the necessary spectral line surveys of its 158-\micron line have not, until recently been available.  The COBE FIRAS instrument made the only large-scale survey of spectrally unresolved  [C\,{\sc ii}] \citep{Wright1991,Bennett1994}, however, COBE with its 7$\degr$ beam and $\sim$1000 \kms velocity resolution, is unable to resolve the latitudinal distribution.  There are two moderate-scale Galactic surveys of spectrally unresolved [C\,{\sc ii}], the Far-Infrared Line Mapper (FILM) onboard the Infrared Telescope in Space (IRTS) \citep[][]{Shibai1994,Makiuti2002} and  the Balloon-borne Infrared Carbon Explorer (BICE) \citep{Nakagawa1998}, and an earlier  small-scale survey with the Balloon-borne Infrared Telescope  (BIRT) \citep{Shibai1991}.  However, there is only one spectrally resolved survey, the  {\it Herschel} open time key program, Galactic Observations of Terahertz C+, hereafter GOT C+ \citep[see][]{Langer2010,Pineda2013,Langer2014}.  FILM and BICE had an angular resolution of order ten to fifteen arcminutes, sufficient to determine the latitudinal $b$ distribution of [C\,{\sc ii}], but the velocity resolution was, at best, $\sim$175 \kms for BICE (for FILM it was $\sim$750 \kms and for BIRT 143 \kmsnospace).  Only the GOT C+ survey had the spectral resolution ($<$1 \kmsnospace) sufficient to resolve the velocity structure of individual gas clouds and thus locate their Galactic radius using a position--velocity rotation curve.  However, GOT C+ surveyed [C\,{\sc ii}] sparsely in longitude, $l$, and latitude, $b$.

Here we determine the average scale height distribution of [C\,{\sc ii}] in $z$ by combining the high spectral resolution [C\,{\sc ii}]  radial distribution from GOT C+ at $b = 0\degr$ with the BICE latitudinal angular distribution in $b$ by adopting a hydrostatic model result for the distribution in $z$.  We use the BICE survey to determine the scale height of [C\,{\sc ii}] emission because it had better coverage in both longitude and latitude in the Galactic disk than the BIRT and FILM surveys. BICE had an angular resolution of 15$^\prime$, and spectral resolution of $\sim$175 \kmsnospace.

The GOT C+ survey contains several hundred lines-of-sight of spectrally resolved [C\,{\sc ii}] emission throughout the Galactic disk from $l=0\degr$ to 360\deg and $b= 0\degr$, $\pm0.5\degr$, and $\pm1.0\degr$.  However, because GOT C+ is a sparse survey it does not have sufficient coverage in latitude $b$ to derive a smooth continuous distribution in the vertical distribution $z$. In contrast, the BICE survey had insufficient spectral resolution, but had much better coverage in $b$, but only observed up to latitudes $b = \pm 3\degr$ and only within longitude $350\degr \le l \le 25\degr$. GOT C+  has a 3-$\sigma$ sensitivity $\sim$ 0.24 K \kms  \citep{Langer2014} which, over the bandwidth of the velocity resolution of BICE, corresponds to 7.4$\times$10$^{-6}$ ergs s$^{-1}$ cm$^{-2}$ sr$^{-1}$. BICE has a  3-$\sigma$ detection limit  $\sim$2$\times$10$^{-5}$ ergs s$^{-1}$ cm$^{-2}$ sr$^{-1}$ \citep{Nakagawa1998}. Thus GOT C+ is almost three times more sensitive than  BICE and capable of detecting the [C\,{\sc ii}] emission seen by BICE.  
 
We also use results from FILM \citep{Shibai1996,Makiuti2002}, which observed [C\,{\sc ii}] at higher latitudes than BICE, to recalibrate the results of \cite{Nakagawa1998} for $|b| = 3\degr$ to $4\degr$. FILM observed [C\, {\sc ii}] along a great circle crossing the plane at $l = 50\degr$  (inner Galaxy) and 230\deg (outer Galaxy), but measured the [C\,{\sc ii}] intensity for larger latitudes than BICE.  However, \cite{Shibai1996} and \cite{Makiuti2002} smoothed their data to 1\deg to improve their sensitivity, insufficient to use at low latitudes to determine the scale height in the disk. 

 \cite{Nakagawa1998} compared the latitudinal distribution of [C\,{\sc ii}] from BICE with those of H\,{\sc i}, $^{12}$CO, and far-infrared dust emission, finding that H\,{\sc i} had the largest distribution in $b$ followed by [C\,{\sc ii}] and then far-infrared dust emission, and that CO had the smallest distribution in $b$.  However, without knowing where the [C\,{\sc ii}] emission came from they could not assign a spatial scale height to the gas traced by the [C\,{\sc ii}] 158-\micron line.   We begin with a summary of the BICE and GOT C+ [C\,{\sc ii}] distributions  and then derive  an approximate relationship between the radial and $b$ distributions that allow us to deconvolve the [C\,{\sc ii}] distribution in $z$ in the disk.  Finally, we compare the [C\,{\sc ii}] scale height with those of CO and H\,{\sc i} and discuss its implications for the understanding the sources of [C\,{\sc ii}] emission.


\section{The z-distribution of [C\,{\sc ii}]}
\label{sec:observations}

\begin{figure}
  \centering
      \includegraphics[width=8.5cm]{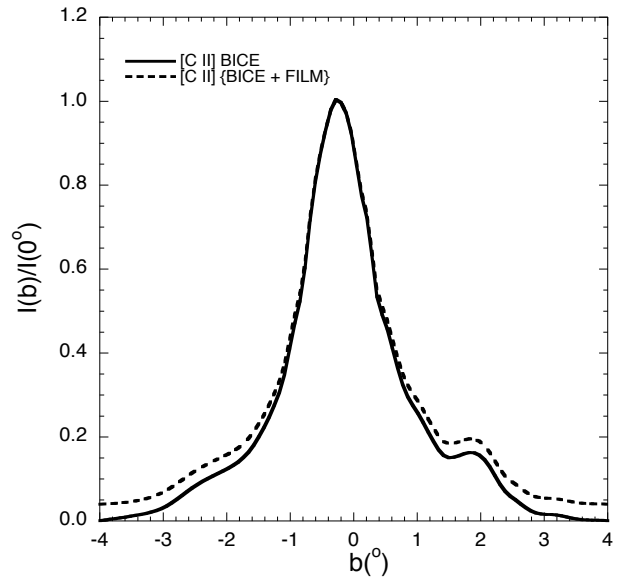}
     \caption{Distribution of the normalized intensity of [C\,{\sc ii}], $I(b)/I(b_c)$, as a function of latitude as derived by \cite{Nakagawa1998} from the BICE survey (solid line).  The modified distribution using the FILM latitudinal data to calibrate the normalized intensity at $b \pm 4\degr$ (dashed line).}
              \label{fig:fig_1_BICE+FILM_FCII_vs_b}
 \end{figure}

To determine the latitudinal distribution of [C\,{\sc ii}]  \cite{Nakagawa1998} averaged the  [C\,{\sc ii}] line intensity (measured in erg s$^{-1}$ cm$^{-2}$ sr$^{-1}$) only over longitudes $5\degr < l <  25\degr$ in order to avoid the Galactic center, and calculated a relative intensity as a function of $b$ such that the peak of this distribution is unity.  Here we define the relative intensity as $I(b)/I(b_c)$, where $I(b)$ is the intensity along $b$, and $I(b_c)$ is the intensity at the peak of the distribution located at $b_c$ (i.e. this term accounts for any offset in the peak from $b=0\degr$). However, \cite{Nakagawa1998}  set the normalized intensity to zero at $b= \pm 4\degr$, thus suppressing potential contributions from $|b|$= $3\degr$ to $4\degr$.  The relative intensity distribution from BICE  is plotted in their Figure 9, and recreated here in  Figure~\ref{fig:fig_1_BICE+FILM_FCII_vs_b}.  \cite{Nakagawa1998} also plotted the corresponding relative intensity of other ISM gas and dust tracers: H\,{\sc i}, $^{12}$CO(1$\rightarrow$0), and far-infrared continuum from IRAS. They derived the full width at half maximum (FWHM) in $b$, for these four tracers and their results are summarized here in Table~\ref{tab:facilities}. In Figure~\ref{fig:fig_1_BICE+FILM_FCII_vs_b} it can be seen that the peak in the [C\,{\sc ii}] relative intensity is shifted slightly below the plane to $b_c \sim -0.2\degr$; the far-IR emission also peaks there (see their Figure 9), but their plot of the CO and H\,{\sc i} peak at $b = 0\degr$. 

The assumption that $I(b = \pm 4\degr)/I(b_c) = 0$ is in conflict with the FILM observations of [C\,{\sc ii}] and could introduce a slight error in determining the FWHM.   \cite{Shibai1996} observed that [C\,{\sc ii}] emission in the FILM  latitudinal scans extended out to perhaps $\pm 50\degr$ (see their Figure 1, which plots the [C\,{\sc ii}] intensity distribution in latitude smoothed over a 1\deg beam). They interpreted the latitudinal [C\,{\sc ii}]  distribution as having two components: (1) concentrated emission in the disk; and, (2) a weaker component, decreasing slowly with $b$ beyond $\sim10\degr$.  \cite{Makiuti2002} analyzed the distribution at large $b$ and conclude that it mainly traces the WIM at high latitudes.  We have fit the wings of the FILM inner Galaxy scan from $4\degr$ to $10\degr$ in order to recalibrate the BICE distribution such that the BICE relative intensity corresponds to the relative intensity ratio observed by FILM at $\pm 4\degr$.  We do not use the FILM fit beyond $\pm 4\degr$ because the FILM observing path results in $b$ being a strong function of longitude $l$ and it does not cover the regions needed to compare GOT C+ with the BICE survey, and because at large $b$ FILM is surveying the Galaxy near the solar radius. The revised BICE  distribution is shown in Figure~\ref{fig:fig_1_BICE+FILM_FCII_vs_b} and labeled BICE + FILM.  This recalibration makes very little difference to the bulk of the [C\,{\sc ii}] distribution, but, as will be seen later, provides a much better fit to the wings at $|b| \ge 2\degr$. 


    \begin{table}
\caption{$b$(FWHM) of gas and dust tracers: \cite{Nakagawa1998}}
\label{tab:facilities}
\begin{tabular}{l c c c c l}
\hline\hline
Tracer & FWHM in $b$  & Data source & \\
\hline
 H\,{\sc i}& 1.96\degr &  \cite{Hartmann1997} \\
${\rm [}$C\,{\sc ii}${\rm ]}$ & 1.32\degr & \cite{Nakagawa1998} \\
far-IR dust & 1.18\degr  &  \cite{Beichman1988} \\
 CO & 1.07\degr  & \cite{Dame1987} \\
\hline
\hline
\end{tabular}\\
\end{table}

\begin{figure}
  \centering
   \includegraphics[width=8.5cm]{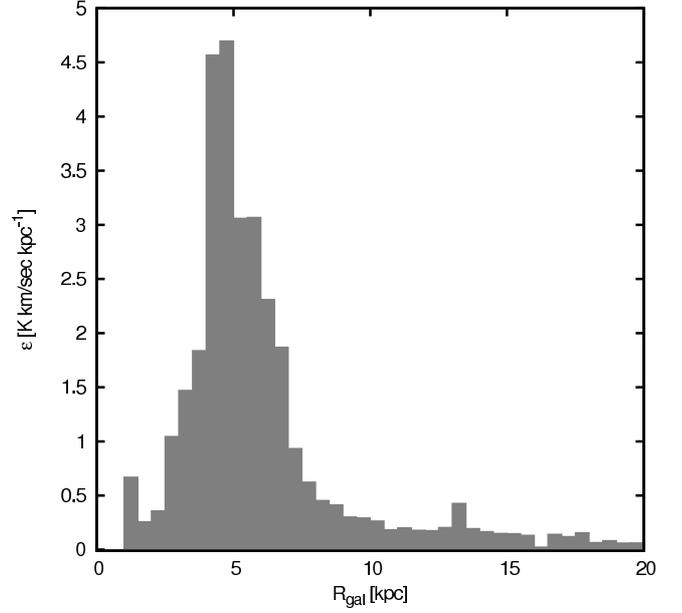}
     \caption{Galactic radial emissivity of [C\,{\sc ii}], $\epsilon_{\rm [C II]}(R_{\rm gal})$, derived by \cite{Pineda2013} using the GOT C+ [C\,{\sc ii}] survey at $b = 0\degr$.}
              \label{fig:fig_2_epsilon_CII_vs_R}
 \end{figure}

The GOT C+ survey observed spectrally resolved [C\,{\sc ii}] along 151 lines-of-sight covering $0\degr < l <360\degr$ at $b=0\degr$.  The details of the observing mode and data reduction are discussed in \cite{Pineda2013}, and representative spectra can be viewed there and in \cite{Langer2014}.    In \cite{Pineda2013} the [C\,{\sc ii}] spatial-velocity maps were used along with a Galactic rotation curve to assign the intensity as a function of Galactic radius.   These intensities were then summed in rings about the Galactic center and then used to calculate the  azimuthally averaged emissivity for [C\,{\sc ii}], $\epsilon_{\rm [CII]}(R_{\rm gal})$ as a function of Galactic radius, $R_{\rm gal}$, in units of K \kms kpc$^{-1}$, except for the innermost 0.5 kpc, because this region is undersampled. In Figure~\ref{fig:fig_2_epsilon_CII_vs_R} we reproduce the radial distribution for $\epsilon_{\rm [CII]}(R_{\rm gal})$ from Figure 7 in  \cite{Pineda2013}.  In Figure~\ref{fig:fig_2_epsilon_CII_vs_R} it can be seen that the [C\,{\sc ii}] emissivity peaks in the molecular ring at about 5 kpc and most of the emission comes from $R_{\rm gal} =$  3 to 7 kpc. 

Our approach assumes that the GOT C+ radial distribution sees all the [C\,{\sc ii}] components observed by BICE including the WIM, which is believed to    more prominent  at higher latitudes than in the plane.  In the Galactic plane for $b$=0$\degr$ the contribution of the WIM component in the radial profile was estimated by \cite{Pineda2013} to be only $\sim$4$\%$ using the electron abundance of low density ionized gas in the plane as given by the NE2001 code \citep{Cordes2002}.  Note that this value is only an estimate and is not directly observed in the GOT C+ data.  However, \cite{Velusamy2012} used the GOT C+ data to detect emission from the compressed WIM along selected lines of sight corresponding to the spiral arm tangencies.  Combined with its higher sensitivity and adequate sampling in the plane, the distribution of [C\,{\sc ii}] emission in the GOT C+ radial profile contains all [C\,{\sc ii}] features as seen by  BICE, including the  emission from the WIM. Therefore we believe that the combined effects of all [C\,{\sc ii}] components will be fully represented in our solution for the z-scale derived below.

We can determine the average $z$ distribution  in the inner Galactic disk by combining the GOT C+ radial distribution of $\epsilon_{\rm [CII]}(R_{\rm gal})$ for $b = 0\degr$ with the modified BICE + FILM latitudinal distribution  if we know the functional form of the emissivity in $z$, and if we assume that the distribution in $z$ is independent of location in the inner Galactic disk.  For the first assumption, models of hydrostatic equilibrium of gas in the ISM suggest a Gaussian distribution for many ISM components, however, the second assumption can only be correct on average, because the pressure in the ISM varies across the Galaxy, and the observed scale heights for CO and H\,{\sc i} vary with Galactic radius.  

\begin{figure}
  \centering
   \includegraphics[width=9.0cm]{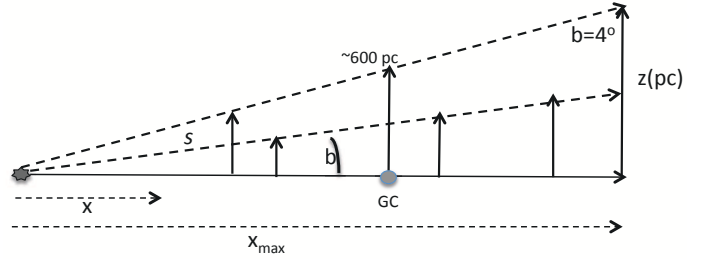}
     \caption{Schematic showing two representative lines of sight (dashed lines) from the solar system through the Galactic disk (GC labels the Galactic center and $b$ labels the latitude).  The intensity of [C\,{\sc ii}] is the integral of the emissivity, $\epsilon$, along the path length $s$ (see text).  
The actual lines of sight used to calculate the intensity cover $l$ = 5$\degr$ to 25\degr.}
              \label{fig:fig_3_CII_LOS_integral}
 \end{figure}

Adopting these two assumptions the averaged [C\,{\sc ii}] intensity observed by BICE + FILM along a line-of-sight is just the integral of the emissivity along a path length $s$ from the solar system, as illustrated in  Figure~\ref{fig:fig_3_CII_LOS_integral} for $l$=0\degr.  If we know the value of the emissivity  at $z = 0$ we can determine its value at height $z$ for a given scale height of the Gaussian distribution.  Thus the intensity measured by BICE + FILM as a function of $b$ is related to the [C\,{\sc ii}] emissivity, $\epsilon_{\rm [CII]}(R_{\rm gal},l,x,z)$, as a function of the distance to the source in the plane, $x$, the height above the plane, $z$, along a line-of-sight longitude  $l$, and latitude, $b$. 
Most of the observed [C\,{\sc ii}] intensity measured by BICE comes from the near side of the inner Galaxy for  $|b| \ge 1\degr$ because at higher latitudes the line of sight passes far above the plane on the far side of the Galaxy.  For example, at  $b = 1\degr$ the line of sight, $s$, passes $\sim$150 pc above the Galactic center, while for $b = 3\degr$,  it is $\sim$450 pc above the plane, and on the far side of the molecular ring it is $>$ 600 pc above $b = 0\degr$.   

We can write the intensity as a function of latitude, $I(b)$, along a given line of sight in terms of the integration of the emissivity as follows,

 \begin{equation}
I_{\rm [CII]}(b)=\int_0^{s_{max}} \epsilon_{\rm [CII]}(s)ds
\end{equation}

\noindent where the integral of the emissivity $\epsilon_{\rm [CII]}(s)$ is along a line $s$ from the solar system, $s_{\rm min}$=0 to a maximum distance $s_{\rm max}$, where the emissivity is small.  As illustrated in Figure~\ref{fig:fig_3_CII_LOS_integral}  we can relate the intensity of emission at each location along $s$ to the emission along the plane at $b = 0\degr$, by changing the integration along $s$ to $x$ in the plane.  Substituting $x = s\cos(b)$ in Equation 1 we get,

\begin{equation}
I_{\rm [CII]}(l,b)=\int_0^{x_{max}} \epsilon_{\rm [CII]}(x )f(z)(cos(b))^{-1} dx
\end{equation}

\noindent where $f(z)$ is the emissivity distribution in $z$. 

We have chosen to integrate up to $x_{\rm max} =$ 28 kpc,  which is on the far side of the Galaxy and where the GOT C+ [C\,{\sc ii}] emissivity is very small.  However, for all practical purposes, as discussed above, most of the 
contribution to the intensity comes from the near side of the Galaxy, except for the very lowest values of $b$. The usual form for $f(z)$ derived from equations of hydrostatic equilibrium balancing the ISM pressure with the gravitational force of the stars and gas in the Galaxy is a Gaussian \cite[c.f.][]{Spitzer1978}, although other dynamical processes (e.g. supernova, outflows) can lead to non-Gaussian terms. Here we assume a Gaussian distribution for the [C\,{\sc ii}] emissivity of the form,

\begin{equation}
f(z) = f(z_c)e^{-0.5((z-z_c)/z_0)^2}
\end{equation}

\noindent where $z_0$ is the scale height, $z_c$ accounts for an offset in the peak of the distribution, and $f(z_c)$ normalizes the distribution to unity.  The FWHM([C\,{\sc ii}]) is equal to 2(2ln2)$^{0.5}$z$_0$. We substitute this form in Equation 2 and, as can be seen in Figure~\ref{fig:fig_3_CII_LOS_integral}, rewrite the integral using $z=x \cos(b)$.  The relative intensity, $F(b)=I(b)/I(b_c)$ is given by,

\begin{equation}  \label{eq:Equation_F(b)}
F(b)=\frac{\int_0^{x_{max}} \epsilon_{\rm [CII]}(x)e^{-0.5((xsinb - z_c)/z_0)^2}cos(b)^{-1}dx}{\int_0^{x_{max}} \epsilon_{\rm [CII]}(x)e^{-0.5(z_c/z_0)^2}dx}.
\end{equation}

\noindent where $\cos(b=0\degr)$ is unity in the denominator.

As mentioned above, \cite{Nakagawa1998} calculated the relative intensity as a function of $b$ by averaging the BICE observations over longitudes from 5\deg to 25\degr, to avoid the Galactic center, thus omitting [C\,{\sc ii}] over a region with Galactic radius $\sim$0.75 kpc.  {We solve for the average scale height covering  longitudes 5\deg to 25$\degr$, where we use the GOT C+ radial profiles to calculate the intensity along $l$ for $b=$0\degr. We also assume that [C\, {\sc ii}] emission is optically thin because most of the GOT C+ [C\,{\sc ii}] spectra have a main beam temperature much less than the kinetic temperature (c.f. discussion of C$^+$ excitation and [C\,{\sc ii}] radiative transfer in \citealt{Goldsmith2012}).  

We iterated on the two parameters, $z_0$ and $z_c$ in Equation~\ref{eq:Equation_F(b)} to minimize the rms deviation of the model averaged over five longitudes compared to the modified BICE + FILM  [C\,{\sc ii}] distribution.  The best fit is given by  $z_0 =$  73 pc and $z_c =$ -28 pc, and is listed in Table~\ref{tab:scaleheight}, along with the corresponding FWHM([C\,{\sc ii}]) = 172 pc.   We also fit the original BICE distribution without the FILM correction and find the same offset and a slightly different FWHM = 170 pc.  The fitting parameters to the BICE and the BICE + FILM distributions do not differ significantly as the  central $\pm 2\degr$ dominates the quality of the fit. However, the fit is much better for the modified BICE + FILM distribution for $|b| > 2\degr$. In Figure~\ref{fig:fig_4_fit_CII_BICE_FILM_vs_b} we show a plot of the  relative intensity for the BICE + FILM  combination and for just the original BICE distribution, compared to the fitted distribution as a function of $b$.  It can be seen that our fit is very good over the complete range $b = -4\degr$ to +$4\degr$, differing by at most $\sim$15\% from the modified BICE + FILM distribution, except at the bumps in the shoulders ($b\sim \pm 2\degr$) where the difference is as much as 30\%. 

\begin{figure}
  \centering
   \includegraphics[width=8.5cm]{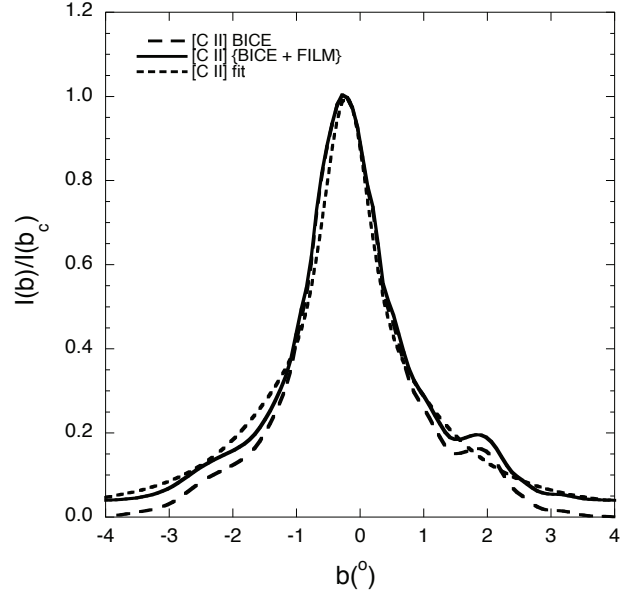}
     \caption{Model fit (dotted line) of the normalized intensity using the GOT C+ observations at $b=0\degr$ compared to the normalized intensity from the modified BICE + FILM profile (solid line) versus latitude $b$.  The unmodified BICE profiles are shown for comparison (dashed line).}
              \label{fig:fig_4_fit_CII_BICE_FILM_vs_b}
 \end{figure}
 
 \cite{Nakagawa1998} calculated the total flux measured by BICE over the entire mapped angular region $|b|\le 3\degr$ and 350\degr$\le l \le$25\degr\, to be 6$\times$10$^{-6}$ ergs s$^{-1}$ cm$^{-2}$.  The BICE calibration errors are $\pm$35$\%$ and include an estimated uniform offset of 2$\times$10$^{-5}$ ergs s$^{-1}$ cm$^{-2}$  sr$^{-1}$, comparable to the calibration uncertainty (see their Section 3.1). FILM observations are better calibrated than those of BICE and \cite{Makiuti2002} compared FILM with BICE and found that the [C\,{\sc ii}] line intensity for FILM is $\sim$65$\%$ of BICE for $I($[C\,{\sc ii}]) $> $10$^{-4}$ ergs s$^{-1}$ cm$^{-2}$  sr$^{-1}$ and $\sim$85$\%$ for $I$([C\,{\sc ii}]) $<$ 10$^{-4}$ ergs s$^{-1}$ cm$^{-2}$  sr$^{-1}$.  To confirm that GOT C+ and BICE are sensitive to the same ISM gas components we calculated the total flux that would be detected by GOT C+ over this same angular area, $|b|\le 3\degr$ and 350\degr$\le l \le$25\degr\,, using the emissivity in the plane ($b = 0\degr$) in \cite{Pineda2013} and the Gaussian distribution in $z$ with a FWHM = 172 pc.  We calculate from GOT C+ a total flux $\sim$4.5$\times$10$^{-6}$ ergs s$^{-1}$ cm$^{-2}$.  This value is about 25$\%$ less than calculated in the BICE survey, however, it is in good agreement if we recalibrate BICE using the calibration from FILM.  The agreement between GOT C+ and BICE is well within the errors of the BICE result, and supports our assumption that GOT C+ and BICE trace the same ISM gas components.
 
The FWHM for CO has been estimated by several authors using different Galactic surveys, including \cite{Sanders1984}, \cite{Dame1987,Dame2001}; \cite{Bronfman1988}; \cite{Clemens1988}, \cite{Malhotra1994}, and \cite{Jackson2006}. They all find about the same result that CO emission peaks between 4 and 7 kpc, and in the Galactic center, and that the scale height increases with increasing Galactic radius beyond 2 kpc and is offset below  $b=0\degr$.  \cite{Sanders1984} found that the FWHM (or 2$z_{\rm 1/2}$ in their notation) ranged from 60 to 140 pc for $R_{\rm gal}$ increasing from 3 to 8 kpc.  To compare the CO distribution with the average derived for [C\,{\sc ii}] we calculated the average FWHM(CO) using their results from 3 to 8 kpc where the [C\,{\sc ii}] is largest. We find  $<$FWHM$>$ = $\sim$110 pc and $z_c \sim$-25 pc, and these values are listed in Table~\ref{tab:scaleheight}.  Their CO offset below the plane is very close to what  we derived for [C\,{\sc ii}], however, the CO FWHM is noticeably smaller than that for [C\,{\sc ii}].

The H\,{\sc i} distribution in the Galaxy has been discussed by many authors \cite[c.f.][]{Boulares1990,Dickey1990,Narayan2002,Kalberla2008,Kalberla2009}. Here we will use the parameters from \cite{Dickey1990} for fitting the distribution of H\,{\sc i} relative intensity with $z$.  H\,{\sc i} is roughly constant from $R_{\rm gal}$ = 4 to 8 kpc, but the distribution in $z$ is complicated by having several components, some of which extend into the halo.  \cite{Dickey1990} find that the best estimate of the $z$ mean density distribution, $n(z)$, is given by two Gaussians with peak mean densities of $n_{\rm 1}(0)$=0.395 and $n_{\rm 2}(0)$=0.107 cm$^{-3}$ and FWHM$_{\rm 1}$= of 212 and FWHM$_{\rm 2}$=530 pc, plus an exponential with a peak mean density, $n_3(0) =$ 0.064 cm$^{-3}$ and scale height 403 pc. In Table~\ref{tab:scaleheight} we list their two Gaussian parameters and  one exponential parameter, but to calculate a relative H\,{\sc i} intensity we normalize their contributions to the total intensity using the corresponding central mean densities to weight the contributions, where the total $F(z) = F_{\rm 1}(z) + F_{\rm 2}(z) + F_{\rm 3}(z)$, with $F_{\rm i}(b_c)=n_{\rm i}(b_c)/n_{\rm tot}(b_c)$, and $n_{\rm tot}(b_c)=n_{\rm 1}(b_c)+n_{\rm 2}(b_c) +n_{\rm 3}(b_c)$.  We have included an offset $z_c = - 25$ pc for H\,{\sc i} in Table~\ref{tab:scaleheight}, to facilitate comparison of the relative distributions of the three gas tracers. The spatial resolution of H\,{\sc i} Galactic surveys is not high enough to resolve clearly an offset $z_c({\rm HI})$ of order 25 pc, but there is some indication that the solar system is offset from the warped Galactic plane as seen in the Leiden-Argentine-Bonn 21-cm survey \cite[see Figure 3 in][]{Kalberla2009}.  

In Figure~\ref{fig:fig_5_CII_CO_HI_vs_z} we plot the Gaussian distributions of the relative  intensities as a function of $z$ for [C\,{\sc ii}], CO, and H\,{\sc i} using the average FWHM for each of the components listed in Table~\ref{tab:scaleheight}.  As can be seen in Figure~\ref{fig:fig_5_CII_CO_HI_vs_z} the average distribution of $^{12}$CO as a function of $z$ in the Galactic disk is narrower than [C\,{\sc ii}], and both are narrower than that for H\,{\sc i}. The distribution for [C\,{\sc ii}] consists mainly of a disk component confined to $\pm 200$ pc.

\begin{figure}
  \centering
   \includegraphics[width=8.5cm]{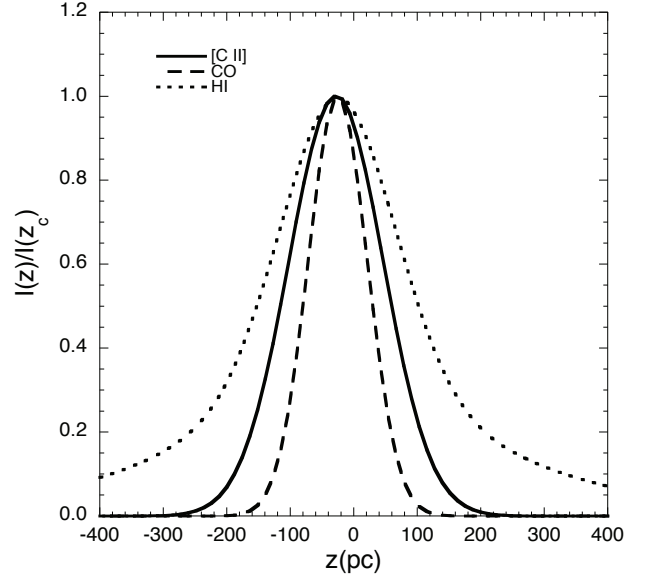}
     \caption{Gaussian fits for the normalized intensities of [C\,{\sc ii}], CO, and H\,{\sc i} as a function of $z$.  The fits include the measured offsets  $z_{\rm offset}$ for CO and [C\,{\sc ii}], and an inferred offset for H\,{\sc i} equal to that for CO.  Towards the inner Galaxy CO has, on average, the narrowest distribution in the disk, followed by [C\,{\sc ii}], and  H\,{\sc i} has the thickest distribution.}
              \label{fig:fig_5_CII_CO_HI_vs_z}
 \end{figure}

    \begin{table}
\caption{Tracer fitting parameters}
\label{tab:scaleheight}
\begin{tabular}{l c c c c  c c c l }
\hline\hline
Tracer & z$_0$ & $<$FWHM$>$ & z$_{\rm c}$ & F($b_c$) & F($\pm$4\degr) & Reference\\
     & (pc) & (pc)   & (pc) & \\
${\rm [}$C\,{\sc ii}${\rm ]}$ & 73 &  172  & -28 & 1.0 & 0.04 & This paper\\
$^{12}$CO & 46.7 &   110   & -25 & 1.0 &  & 1  \\
H\,{\sc i}&  &  212 & -25$^5$ & 0.698 & & 2 \\
 H\,{\sc i}&  & 530 & -25$^5$ & 0.189 & & 3 \\
H\,{\sc i}&  & 403 & -25$^5$ & 0.113 &  & 4 \\ 
\hline
\hline
\end{tabular}\\
1) Derived from results in \cite{Sanders1984} averaged over the range, $R_{\rm gal}$ = 3 to 8 kpc.;  (2)  \cite{Dickey1990}  Gaussian component 1. (3) \cite{Dickey1990}  Gaussian component 2. (4) \cite{Dickey1990} exponential component.  (5) Assumed offset in H\,{\sc i} to facilitate comparison of the distributions.  
\end{table}


\section{Discussion}
\label{sec:discussion}

There is a very simple explanation why the scale height for [C\,{\sc ii}] is more than that of CO but less than that of H\,{\sc i}.  While H\,{\sc i} traces mainly the atomic hydrogen clouds in the Galaxy and CO traces the dense molecular clouds, [C\,{\sc ii}] traces both of these regions, as well as the diffuse molecular clouds that have [C\,{\sc ii}] but no CO emission (CO-dark H$_2$ clouds), and the WIM.  The solution for the density distribution of clouds in hydrostatic equilibrium is a Gaussian function, $\propto exp(-0.5(z/z_0)^2)$, with the scale factor, $z_0$ proportional to the velocity dispersion, $<v_i^2>^{0.5}$, where $i$ labels the ISM cloud component. The diffuse atomic clouds with their lower densities have higher velocity dispersions, $<v_{\rm H}^2>^{0.5}$  $\sim 10$ \kmsnospace, while those for the denser CO clouds have  $<v_{\rm CO}^2>^{0.5}$ $\sim 5$ \kms \cite[see discussion and models in][]{Narayan2002}. Thus in hydrostatic equilibrium, for an equal ISM pressure, the H\,{\sc i} clouds have a larger scale height than those of the denser CO clouds. 

\cite{Pineda2013} found that the PDRs of dense molecular clouds emit $\sim$43$\%$ of the total [C\,{\sc ii}] throughout the plane at $b$ = 0\degr, diffuse atomic hydrogen clouds $\sim$23$\%$, diffuse molecular clouds (CO-dark H$_2$ clouds)  $\sim$30$\%$., and for the WIM estimated $\sim$4$\%$.  In the inner Galaxy, $R_{\rm gal} <$ 9 kpc,  where we evaluate the [C\,{\sc ii}] distribution, these percentages are only slightly different. Therefore, it is no surprise that the distribution of [C\,{\sc ii}], which arises from all ISM components, would have a distribution intermediate between that of the dense CO clouds and less dense atomic H\,{\sc i} clouds and the WIM.  Thus [C\,{\sc ii}] traces a mixture of clouds of different mass, density, and velocity dispersion, some of which are also traced by H\,{\sc i} and CO, and the WIM. The different scale heights then depend on the different physical properties and energetics of the clouds that enter into the hydrostatic mechanisms responsible for the distribution of gas in the plane. 

As seen in Figure~\ref{fig:fig_5_CII_CO_HI_vs_z}, the derived $z$ distribution in [C\,{\sc ii}] does not follow that of  H\,{\sc i} for $z$ greater than about $\sim$200 pc.  At this height there are few dense molecular clouds as traced by CO and likely very few diffuse molecular clouds (CO-dark H$_2$ clouds), so any [C\,{\sc ii}] would have to come from the H\,{\sc i} clouds and$/$or the WIM.   \cite{Makiuti2002} compared the distribution of [C\,{\sc ii}] and H\,{\sc i} at high  latitudes and conclude that the [C\,{\sc ii}] emission comes primarily from the WIM (see their Figure 3).  However, the FILM results at high latitude are limited to a region near the solar radius and cannot be extrapolated across the Galaxy.  The combined GOT C+ and BICE data suggest that this conclusion also holds for the inner Galaxy as well, because H\,{\sc i} clouds high above the plane have low densities and are not likely to emit [C\,{\sc ii}] efficiently.  This low emissivity is due to the difference in the excitation conditions for H\,{\sc i} and [C\,{\sc ii}]. The intensity of H\,{\sc i} is proportional to its column density in the optically thin regime, 

\begin{equation}
I([{\rm H\,I}])  = 5.5\times 10^{-19}N({\rm H\,I}),
\end{equation}

\noindent in units of (K \kmsnospace), and is relatively insensitive to density and kinetic temperature. In contrast the  [C\,{\sc ii}] emission is very sensitive to kinetic temperature, $T_{\rm kin}$ because the energy of the upper level $^{\rm 2}P_{\rm 3/2}$, $E_u/k = 91.25 K$ is typically higher than the gas temperature in neutral clouds, and density where the atomic, n(H), and molecular, n(H$_2$), hydrogen densities are much lower than the critical densities for thermalizing the C$^+$,  $n_{\rm cr}(H)\sim$ 3000 cm$^{-3}$ and  the recently revised value $n_{\rm cr}(H_2)\sim$4500 cm$^{-3}$ \citep{Wiesenfeld2014}.  The radiative transfer equation for the [C\,{\sc ii}] intensity for optically thin emission is given in \cite{Goldsmith2012} and \cite{Langer2014}, and, in the diffuse clouds, such that the intensity can be written as,

\begin{equation}
I_j([{\rm C\,II}])  = 1.73\times10^{-16}  e^{{-\Delta}E/kT} \frac{ n(j)}{n_{\rm cr}(j)} N_j({\rm C^+}) ,
\end{equation}

\noindent where $I$ is in units of K \kms and the index $j$ labels H\,{\sc i} or H$_2$.  Therefore, while the H\,{\sc i} intensity depends only on the column density of atomic hydrogen, the [C\,{\sc ii}] intensity also depends very sensitively on the density and will be much smaller in low density atomic hydrogen clouds above the plane.  

The scale height for [C\,{\sc ii}] derived here depends on the radial distribution derived from the GOT C+ sampling at $b$=0$\degr$, which as noted above is a sparse sample.  The premise of the GOT C+ survey was that a well designed unbiased sampling in longitude would represent statistically the distribution of [C\,{\sc ii}] in the Galactic plane.  Therefore, the fact that GOT C+, along with our model of the $z$ distribution, reproduces the total flux observed by BICE (rescaled to the FILM calibration) supports this approach.

Another potential uncertainty is the adoption of a Galactic rotation curve in \cite{Pineda2013} to locate the source of the [C\,{\sc ii}] emission.  In \cite{Langer2014} we adopted a rotation curve based on gas-flow hydrodynamical models to assign a distance based on velocity.  We found that it made a difference mainly in the inner Galaxy, $|l|\le$6$\degr$, but this region is mostly excluded from the BICE analysis (see above).  There are also regions where clouds have peculiar velocities due to Galactic dynamics, where the rotation curve may assign the wrong distance.  For example, \cite{Zhang2014} find non-rotational cloud motions at the end of the Galactic bar from parallax observations of masers at $l\sim$30\degr. While 30\degr\ is outside of the longitudinal range observed by BICE, this region contributes to our GOT C+ data set.   We cannot quantitatively assess the error introduced into our radial distribution but note that, because we average lines of sight from all across the Galaxy, the edges of the bar contribute a small fraction of the emission in any given ring.

We cannot calculate the radial dependence for the [C\,{\sc ii}] FWHM from the GOT C+ survey without a better sampling in $b$.  However, to gain some insight on the effect of a variable FWHM($R_{gal}$), we assume that it varies similar to that for CO.  \cite{Sanders1984} and \cite{Clemens1988} found that the CO scale height varied roughly as $R_{\rm gal}^{0.5}$  between 3 and 9 kpc.  We replaced $z_o$ in Equation (3) with one that varied $\propto R_{\rm gal}^{0.5}$ for $R_{gal} >$3 kpc at the value for 3 kpc.   We solved for the scale factor that best fit the BICE distribution in $b$, similar to what was done to determine an average scale factor.  We find that the best fit is given by, FWHM($R_{\rm gal})=172(R_{gal}/4.7)^{0.5}$ pc.  Thus the average value for FWHM for [C\,{\sc ii}] of 172 pc corresponds to the radial solution at $\sim$ 4 to 5 kpc, essentially in the molecular ring. For the assumed radial dependence, the FWHM ranges from $\sim$140 pc to $\sim$ 230 pc over Galactic radii 3 kpc to 8 kpc. 



\section{Summary}
\label{sec:summary}

We have combined the  GOT C+ spectrally resolved [C\,{\sc ii}] survey in the Galactic plane at $b = 0\degr$ with the latitudinal distribution derived from the BICE survey of spectrally unresolved [C\,{\sc ii}] to derive, for the first time, the average scale height of [C\,{\sc ii}] over the inner Galactic plane. GOT C+ is slightly more sensitive than BICE and the total flux measured by GOT C+ is close to that of BICE given the uncertainties of the BICE calibration.  Therefore these two surveys are likely tracing the same ISM gas components.  The average distribution in the inner Galactic disk is well fit by a single Gaussian with FWHM([C\,{\sc ii}]) = 172 pc and an offset -28 pc below the plane ($b = 0\degr$). 

In this paper we find that the [C\,{\sc ii}]  distribution is larger in $z$ than that of CO, but smaller than H\,{\sc i}.  The origin of the [C\,{\sc ii}] emission has been attributed to different sources by various authors based on the spectrally unresolved surveys. However, the result here suggests to us  a more complicated picture with [C\,{\sc ii}] tracing a mix of ISM cloud categories. The GOT C+ data for $b \ne 0\degr$ may be able to give some insights on the distribution of the different ISM components, but to determine completely the distribution of [C\,{\sc ii}] for the separate ISM components as a function of Galactic radius and $z$, we will need more detailed spectrally resolved latitudinal maps across the Galaxy and with finer steps in $b$. We also need to extend the spectral line observations to higher values of $b$ than observed in the GOT C+ survey to understand the contributions of the warm ionized medium and low density high latitude H\,{\sc i} clouds to the [C\,{\sc ii}] emissivity above the disk.


\begin{acknowledgements}
We thank the referee for a careful reading of the manuscript and several suggestions that improved the discussion. We also thank P. F. Goldsmith for constructive comments and edits. This work was performed at the Jet Propulsion Laboratory, California Institute of Technology, under contract with the National Aeronautics and Space Administration.   
 
\end{acknowledgements}


\bibliographystyle{aa}
\bibliography{aa_2013_23281_Langer_refs}


\end{document}